\documentclass[preprint,aps,amsmath,nofootinbib,tightenlines]{revtex4}
\usepackage{graphicx}
\usepackage{bm}
\usepackage{epsfig}
\usepackage{graphics}

\newif\ifpdf
\ifx\pdfoutput\undefined
\pdffalse % we are not running PDFLaTeX
\else
\pdfoutput=1 % we are running PDFLaTeX
\pdftrue
\fi

\def\itbf#1{\mbox{\boldmath $#1$}}

\def\Dsl{\hbox{/\kern-.6000em D}} %roman D

\def\dsl{\,\raise.15ex\hbox{/}\mkern-13.5mu D}

\def\psip#1{\psi_{\mathbf{#1}}}
\def\chip#1{\chi_{\mathbf{#1}}}

\def\ltap{\ \raise.3ex\hbox{$<$\kern-.75em\lower1ex\hbox{$\sim$}}\ }
\def\gtap{\ \raise.3ex\hbox{$>$\kern-.75em\lower1ex\hbox{$\sim$}}\ }
\def\OMIT#1{}

\def\lsim{\mathrel{\raise.3ex\hbox{$<$\kern-.75em\lower1ex\hbox{$\sim$}}}}
\def\gsim{\mathrel{\raise.3ex\hbox{$>$\kern-.75em\lower1ex\hbox{$\sim$}}}}

\def\msb{{\overline{\rm MS}}}

\newcommand{\bmk}{\mathbf k}
\newcommand{\bmp}{\mathbf p}
\newcommand{\bbmp}{\mbox{\scriptsize\boldmath $p$}}

\newcommand{\bmA}{\mathbf A}

\newcommand{\bmD}{\mathbf D}

\catcode`\@=11
\def\slash{\mathpalette\make@slash}
\def\make@slash#1#2{\setbox\z@\hbox{$#1#2$}%
  \hbox to 0pt{\hss$#1/$\hss\kern-\wd0}\box0}
\catcode`\@=12 % @ signs are no longer letters
\begin{document}
%%%%%%%%%%%%%%%%%%%%%%%%%%%%%%%%%%%%%%%%%%
%Some more stuff to get graphics to work
\ifpdf
\DeclareGraphicsExtensions{.pdf, .jpg}
\else
\DeclareGraphicsExtensions{.eps, .jpg}
\fi
%%%%%%%%%%%%%%%%%%%%%%%%%%%%%%%%%%%%%%%%%%

%%%%%%%%%%%%%%%%%%%%%%%%%%%%%%%%%%%%%%%%%%
%Define Title, Author, Address, Preprint#

\preprint{ \vbox{ \hbox{MPP-2004-171} 
%\hbox{hep-ph/0412258}  
}}

\title{\phantom{x}\vspace{0.5cm} 
Electroweak Absorptive Parts in NRQCD Matching Conditions
\vspace{1.0cm} }

\author{Andr\'e~H.~Hoang and Christoph~J.~Rei\ss er\vspace{0.5cm}}
\affiliation{Max-Planck-Institut f\"ur Physik\\
(Werner-Heisenberg-Institut), \\
F\"ohringer Ring 6,\\
80805 M\"unchen, Germany\vspace{1cm}
\footnote{Electronic address: ahoang@mppmu.mpg.de, reisser@mppmu.mpg.de}\vspace{1cm}}

%\date{\today\\ \vspace{1cm} }

%%%%%%%%%%%%%%%%%%%%%%%%%%%%%%%%%%%%%%%%%%
\begin{abstract}
\vspace{0.5cm}
\setlength\baselineskip{18pt}
Electroweak corrections associated with the instability of the top quark 
to the next-to-next-to-leading logarithmic (NNLL) total top pair threshold
cross section in $e^+e^-$ annihilation are determined. Our method is based on  
absorptive parts in electroweak matching conditions of the NRQCD
operators and the optical theorem. The corrections lead to ultraviolet
phase space divergences that have to be renormalized and lead to NLL mixing
effects. Numerically, the corrections can amount to several percent and are
comparable to the known NNLL QCD corrections.  
\end{abstract}
% \pacs{12.39.Hg,11.10.St,12.38.Bx}
\maketitle

%%%%%%%%%%%%%%%%%%%%%%%%%%%%%%%%%%%%%%%%%%

% \tighten
\newpage
%%%%%%%%%%%%%%%%%%%%%%%%%%%%%%%%%%%%%%%%%%
%Main body of the paper
%\setlength\baselineskip{15pt}

%
%
%
\section{Introduction}
\label{sectionintroduction}

The line-shape scan of the threshold top pair production cross
section $\sigma(e^+e^-\to \gamma^*,Z^*\to t\bar t)$ constitutes a major part
of the top quark physics program at the 
International Linear Collider (ILC) project that is currently being
initiated. Because in the Standard Model the top quark width
$\Gamma_t\approx 1.5$~GeV is much larger than the typical hadronization energy
$\Lambda_{\rm QCD}$, it is expected that the line-shape of the total cross
section is a smooth function of the c.\,m.\,\,energy, and that non-perturbative
effects are strongly suppressed\,\cite{Kuehn1,Fadin1}. The determination of
the top quark mass (in a threshold mass scheme~\cite{Habilitation,CKMworkshop})
is the most important measurement that can be obtained from the threshold
scan since an uncertainty of only around 100~MeV is 
expected\,\cite{TTbarsim,synopsis}. This prospect is quite 
robust, from the theoretical as well as from the experimental point of view,
since it relies mostly on the determination of the c.\,m.\,\,energy where 
the cross section rises. Because the $t\bar t$ pair is produced predominantly
in an S-wave state the rise of the cross section is quite rapid and easily
measurable even in the presence of beam effects~\cite{TTbarsim}.
In addition it will also be possible to determine the
strong coupling $\alpha_s$, the total top quark width $\Gamma_t$ and, if the
Higgs boson is light, the top Yukawa coupling $g_{\rm tth}$. However, the
latter measurements are sensitive to the form and the normalization of the
line-shape. Since the observable cross section is a convolution of the theory
prediction with the partly machine-dependent luminosity
spectrum arising from QED effects~\cite{TTbarsim,Cinabro1}, 
high demands are imposed on theoretical predictions and experimental
analyses to make these measurements possible. In particular, theoretical
predictions need to have a precision at the level of only a few percent.

The common theoretical tool to make computations for top threshold
observables is non-relativistic QCD (NRQCD).\footnote{We use the term
NRQCD to refer to a generic low-energy effective theory which describes
nonrelativistic $t\bar t$ pairs and bound state effects and not for a theory 
valid only for scales $m_t > \mu > m_t v$. For the presentation 
we use the conventions and notations of 
vNRQCD established in~\cite{LMR,HoangStewartultra}, but we emphasize that our
results are generally true.}   
It provides an economic and systematic treatment for the 
non-relativistic expansion and for QCD radiative corrections coming from high
and low 
energies. It has also become evident that it is advantageous to use 
renormalization group methods~\cite{LMR} to resum logarithms of the
top quark velocity $v$ to all orders of QCD perturbation theory in order to 
avoid large normalization uncertainties of at least $20$~\% that are obtained
in fixed-order predictions~\cite{synopsis}. Concerning QCD effects at
next-to-leading-logarithmic (NLL) order for the total cross section only the
$t\bar t$ production current has a non-trivial running, which is fully
known\,\cite{LMR,HoangStewartultra,Manohar1,Pineda1}. At
next-to-next-to-leading-logarithmic (NNLL) order for the total cross section
the QCD evolution of almost all required couplings 
is known~\cite{HoangStewartultra,Pineda2,Hoang3loop,Peninc1} except for
missing  
subleading mixing effects in the running of the heavy quark pair production
current. Theoretical analyses for the total cross section at NNLL
order were given in~\cite{hmst,hmst1,HoangEpiphany}. Currently the
normalization uncertainties of the total cross section from QCD effects are
estimated to be around $6\%$~\cite{HoangEpiphany}. 

While the major focus in recent work was directed on a better understanding of
QCD corrections at NNLL order, a systematic treatment of electroweak
effects at the same level has not yet been accomplished. Electroweak
corrections are responsible for a variety 
of different physical effects. At leading order, the three basic electroweak
effects are the $e^+e^-$ annihilation process that leads to top pair
production by virtual photon and Z exchange, the finite top lifetime, which
can be implemented by the replacement rule $E\to E+i\Gamma_t$~\cite{Fadin1}
($E=\sqrt{s}-2m_t$ being the c.\,m.\,\,energy with respect to the top threshold
and $\Gamma_t$ being the  on-shell top quark width), and the luminosity
spectrum mentioned above. All three effects have so far been treated independently.
It has become the convention that only the first two
effects are included into theoretical predictions while the luminosity
spectrum is accounted for in the experimental simulations~\cite{TTbarsim}. 

At the subleading level a
coherent treatment of electroweak effects has not yet been achieved, although
previous partial analyses have indicated that they can reach the level of
a few percent~\cite{GuthKuehn,HoangTeubnerdist}. It is also evident that at
the subleading level an independent treatment of various different
electroweak effects will, eventually, be impossible.
Since a systematic treatment relies on the consistent
separation of off-shell (non-resonant) and close-to-mass-shell (resonant)
fluctuations, the concept of effective theories appears again to be a highly
efficient tool to make progress~\cite{Beneke1}. For the $t\bar t$ threshold such a
framework is already provided by the effective theory NRQCD itself, since it
automatically achieves an expansion in the off-shellness of the top quark  
through the non-relativistic expansion in $v$. The NRQCD effective theory
formalism can be extended to account for
electroweak corrections. For example, including the electroweak radiative
corrections to the top quark two-point function in the NRQCD matching
conditions one obtains the additional heavy quark bilinear terms\footnote{
Note that in the effective theory we use positive energy spinors for the
antiparticles. 
} 
\begin{eqnarray} \label{Gtop}
 \delta {\cal L} = \sum_{\bbmp} \psip{\bbmp}^\dagger \: \frac{i}{2} \Gamma_t\: 
  \psip{\bbmp} \ +\  
  \sum_{\bbmp}   \chip{\bbmp}^\dagger\: \frac{i}{2} \Gamma_t\: \chip{\bbmp} \,
\end{eqnarray}
in the effective Lagrangian, where  $\psip{p}$ and  $\chip{p}$ represent Pauli
spinor field operators destroying top and antitop quarks, respectively. For
simplicity color indices are suppressed throughout this paper. The terms in
Eq.\,(\ref{Gtop}) reproduce the replacement rule $E\to E+i\Gamma_t$. They
render the effective theory Lagrangian non-hermitian because 
they describe an absorptive on-shell process that has been integrated out from
the theory. Nevertheless, they allow for a correct
determination of the total cross section from the forward scattering
amplitude using the optical theorem and the unitarity of the underlying
theory. In fact, this issue is analogous to the so-called strong phases in QCD
amplitudes that are the basis for the search for CKM CP-violation in a number
of B meson decays~\cite{BaBarBook}.

In this paper we extend this approach 
and investigate the role of absorptive parts related to the top quark decay in
electroweak loop corrections to the NRQCD matching conditions that contribute
to the NNLL total cross section.\footnote{
  In a complete treatment of all
  electroweak effects one can integrate out all electroweak effects associated
  with the massive W, Z and Higgs bosons at the scale $m_t$. Below the scale
  $m_t$ gluons, photons and quarks remain as dynamical degrees of freedom. 
}
For simplicity we neglect the bottom quark mass and set
$V_{tb}=1$, and approximate the W boson and bottom quark as stable
particles.
We demonstrate that these electroweak corrections properly account
for the interference of the dominant double-resonant process $e^+e^-\to t\bar
t\to bW^+\bar b W^-$ with the $v^2$-suppressed single-resonant amplitudes for  
$e^+e^-\to b W^+\bar t\to bW^+\bar b W^-$ and 
$e^+e^-\to t\bar b W^-\to bW^+\bar b W^-$. We also show that absorptive parts
that do not contribute to the $bW^+\bar b W^-$ final state, and are therefore
also not accounted for in line-shape measurements, can be excluded in a
gauge invariant way.  
This requires to include also fields for the electrons and positrons from the
initial state into the effective theory, which act like classic fields for QCD
interactions. The new corrections show interesting features. They slightly
modify the form of the cross section line-shape and lead to UV phase
space divergences
that are directly related to the 
fact that the top quarks are unstable. The divergences lead to anomalous
dimensions of $(e^+e^-)(e^+e^-)$ operators that also contribute to the
absorptive part of the forward scattering amplitude. In the total cross
section these mixing effects contribute at NLL order and represent a novel 
NLL effect that has remained unnoticed in previous work. 
Numerically, the size of the new corrections ranges up to 5\% and partly
compensates for the large QCD corrections found recently in~\cite{Hoang3loop}.

The program of this paper is as follows. In
Sec.\,\ref{sectionpowercounting} we discuss the power counting for the
total cross section with respect to electroweak effects associated with
the top quark decay. We introduce the $(e^+e^-)(t\,\bar t)$ effective
theory operators needed to account for the electroweak absorptive parts
that arise in top pair production or annihilation. In particular, we
show that up to NNLL order there are no contributions from
interference effects originating from (ultrasoft) gluons carrying
momenta of order $m_t v^2$. In Sec.\,\ref{sectionmatchingconditions}
the electroweak absorptive
parts of the matching conditions for the $(e^+e^-)(t \,\bar t)$
effective theory operators relevant for the $bW^+\bar bW^-$ final state
are computed. In
Sec.\,\ref{sectionrenormalization} the resulting NNLL corrections for the
total cross section are determined and the renormalization of the
$(e^+e^-)(e^+e^-)$ effective theory operators needed to account for the
phase space divergences is discussed. In particular, we compute and
solve the anomalous dimensions and determine their contribution to the
total cross section. Sec.\,\ref{sectionanalysis} contains a brief
numerical analysis and in Sec.\,\ref{sectionconclusion} we conclude.

\section{Power Counting and Matching Conditions} 
\label{sectionpowercounting}

In this work we are interested in the total cross section and not in any
differential information on the top decay final states. We therefore include all
effects related to the top quark decay as non-hermitian matching conditions of
effective theory operators that describe the non-relativistic top and antitop 
dynamics and their interactions with soft and ultrasoft gluons. 
We employ gauge invariant operators, and the matching conditions are computed
for on-shell external lines. This allows to maintain gauge invariance in a
transparent way.  

To illustrate the power counting needed to classify the order at which 
these electroweak effects can contribute let us recall the matching conditions
for the bilinear quark field operators. They are obtained by matching top or
antitop 2-point functions in the effective theory to those in the full
electroweak and QCD theory. The result up to NNLL order
reads
%~\cite{HoangTeubnerdist} 
\begin{eqnarray} \label{Lke}
 {\mathcal L}_{\rm bilinear}(x) &=& \sum_{\bbmp}
   \psip{\bbmp}^\dagger(x)   \biggl\{ i D^0 - {(\itbf{p}-i\bmD)^2 \over 2 m_t} 
   +\frac{{\itbf{p}}^4}{8m_t^3}  
   + \frac{i}{2} \Gamma_t \bigg( 1 - \frac{{\itbf{p}}^2}{2 m_t^2} \bigg) 
   - \delta m_t \biggr\} \psip{\bbmp}(x) 
\nonumber\\ & & \qquad
+ (\psip{\bbmp} \to\chip{\bbmp})\,, 
\end{eqnarray}
and includes the terms shown in Eq.\,(\ref{Gtop}). Here, the fields 
$\psip{\bbmp}$ and $\chip{\bbmp}$ destroy top and antitop quarks with
momentum ${\itbf{p}}$, $D^\mu=(D^0,-\bmD)=\partial^\mu + i g A^\mu$ is the
ultrasoft gauge covariant derivative and $\Gamma_t$ is the top quark
width defined at the top quark pole. At order 
$g^2$, $g$ being the SU(2) gauge coupling, the width has the form
\begin{eqnarray} \label{Gamma}
\Gamma_t & = &
\frac{\alpha |V_{tb}|^2 m_t}{16 s_w^2 x}\,(1-x)^2(1+2x)
\,,
\end{eqnarray}  
where $s_w$ ($c_w$) is the sine (cosine) of the weak mixing angle, $\alpha$
the fine structure constant and 
\begin{eqnarray}
x\equiv \frac{M_W^2}{m_t^2}
\,.
\end{eqnarray}
%The term $\delta m_t$ is a residual mass term that arises if a threshold mass
%scheme~\cite{synopsis} is used. 
We use the usual $v$-counting
$D^0\sim m_t v^2\sim\Gamma_t\sim m_t g^2$, which leads to the
scaling relation
\begin{eqnarray}
v\,\sim\,\alpha_s\,\sim\,g\,\sim\,g^\prime
\end{eqnarray}
for the SU(2) and U(1) gauge couplings $g$ and $g^\prime$. Because the weak
mixing is of order one we apply the same counting to the SU(2) and U(1) gauge
couplings. 
The term $\delta m_t$ is a residual mass term of order $v^2$ that arises if a
threshold mass scheme~\cite{synopsis} is used. The LL order terms in
Eq.\,(\ref{Lke}) lead to the top/antitop propagator 
\begin{eqnarray}
  \frac{i}{p^0 - {\itbf{p}}^2/(2m_t) + i\Gamma_t/2 -\delta m_t} \,.
\label{toppropagator}
\end{eqnarray}
Although the exchange of time-like ultrasoft $A^0$ gluons contributes at LL
order according to the $v$-counting, their contribution at LL order
can be removed from the particle-antiparticle sector of the theory
by a redefinition of the top and antitop fields related
to static Wilson lines~\cite{Korchemsky1,Bauer1}.\footnote{  
In an explicit computation, quark pair production and quark-antiquark
scattering diagrams involving the time-like gluons
cancel at LL order. In Coulomb gauge all leading order diagrams with time-like
gluons are dimensionless and individually zero in dimensional
regularization~\cite{LMR}.}  The time dilatation term $\propto
\Gamma_t(\bmp^2/2m_t^2)$ originates from the 
momentum-dependence of the full theory spinors and contributes at NNLL
order.

The $t\bar t$ pair is produced by an electroweak process. As long as
electroweak effects are only treated at leading order in the $v$-expansion it
is sufficient to 
describe $t\bar t$ production by a bilinear quark-antiquark current. 
However, as shown below it is necessary to include the initial-state
$(e^+e^-)$ fields to ensure electroweak gauge invariance at subleading
order in the $v$-expansion. The dominant 
operators that have to be used describing $t\bar t$ spin-triplet production
have the form 
\begin{eqnarray}
{\cal O}_{V,\bbmp} & = & 
\big[\,\bar e\,\gamma_j\,e\,\big]\,{\cal O}_{\bbmp,1}^j
%\Big[\,\bar v_{e^+}(k^\prime)\,\gamma_j\, u_{e^-}(k)\,\Big]\,
%\Big[\,\psi_{\bbmp}^\dagger\, \sigma_j (i\sigma_2)
%  \chi_{-\bbmp}^*\,\Big]
\,,
\\[2mm]
{\cal O}_{A,\bbmp} & = &
\big[\,\bar e\,\gamma_j\,\gamma_5\,e\,\big]\,{\cal O}_{\bbmp,1}^j
%\Big[\,\bar v_{e^+}(k^\prime)\,\gamma_j\,\gamma_5\, u_{e^-}(k)\,\Big]\,
%\Big[\,\psi_{\bbmp}^\dagger\, \sigma_j (i\sigma_2) \chi_{-\bbmp}^*\,\Big]
\,,
\end{eqnarray}
where
\begin{eqnarray}
{\cal O}_{\bbmp,1}^j & = &
\Big[\,\psi_{\bbmp}^\dagger\, \sigma_j (i\sigma_2)
  \chi_{-\bbmp}^*\,\Big]
\,.
\end{eqnarray}
They give the contribution 
$\Delta {\cal L} = \sum_{\bbmp} \left(C_V {\cal O}_{V,\bbmp} +  C_A
{\cal O}_{A,\bbmp} \right) + \mbox{h.c.}$ to the effective theory Lagrangian
where the hermitian conjugation is referring to the operators only. The index
$j=1,2,3$ is summed.  
The corresponding operators describing $t\bar t$ annihilation
are obtained by the hermitian conjugation. Since in this work we only focus on the
electroweak effects related to the top quark decay and, in
particular, neglect QED radiative corrections (including QED binding and the
beam effects mentioned above) the electron and positron fields act like classic
fields in the effective theory.\footnote{For a treatment of
nonrelativistic QED effects within vNRQED see Ref.~\cite{ManoharQED}.} 
In a more complete treatment of electroweak
effects, however, their interactions with photons have to be accounted
for~\cite{Beneke1}. The leading order matching conditions of the operators 
${\cal O}_{V/A,\bbmp}$ obtained from the full theory Born diagrams with photon and Z
exchange are of order $g^2$ and read 
\begin{eqnarray}
C_V^{\rm born}(\nu=1) & = & 
\frac{\alpha\pi}{m_t^2(4c_w^2-x)}\,\bigg[\,
Q_e Q_t(4-x) + Q_t-Q_e-\frac{1}{4s_w^2}
\,\bigg]
\,,
\label{Cvborn}
\\[2mm]
C_A^{\rm born}(\nu=1) & = &
-\,\frac{\alpha\pi}{m_t^2(4c_w^2-x)}\,\bigg[\,
Q_t-\frac{1}{4s_w^2}
\,\bigg]
\,,
\label{Caborn}
\end{eqnarray}
%where $\alpha$ is the electromagnetic coupling at the scale $m_t$,
%$s_w$ and $c_w$ are the sine and cosine of the Weinberg mixing angle and
%\begin{eqnarray}
%x\equiv \frac{M_W^2}{m_t^2}
%\,.
%\end{eqnarray}
where $\nu$ is the vNRQCD renormalization scaling parameter.\footnote{
In vNRQCD the renormalization scales for soft and ultrasoft
fluctuations, $\mu_S$ and $\mu_U$, are correlated through the heavy
quark equation of motion, $\mu_U=\mu_S^2/m_t$. The correlated running
from the hard scale
down to the soft and ultrasoft scales is described by the
dimensionless scaling parameter $\nu$ defined by $\mu_S=m_t\nu$ and
$\mu_U=m_t\nu^2$. Thus $\nu=1$ corresponds to the hard matching scale.
} 
Except for the QED beam effects, which we will not consider here, the
electroweak effects in Eqs.\,(\ref{Gamma},\ref{Cvborn},\ref{Caborn}) 
are the only ones at leading order.
In particular, at this order there are no electroweak effects contributing to
the Coulomb potential acting between the $t\bar t$ pair, 
\begin{eqnarray}
{\cal L}_{\rm pot} & = &
-\sum_{\bbmp,\bbmp^\prime} \frac{{\cal V}_c^{(s)}(\nu)}{(\bmp-\bmp^\prime)^2}\,
\psip{\bbmp^\prime}^\dagger \psip{\bbmp} \chip{-\bbmp^\prime}^\dagger\chip{-\bbmp}
\,,
\label{Lpot}
\end{eqnarray} 
where ${\cal V}_c^{(s)}(\nu)=-4\pi C_F \alpha_s(m_t\nu)$ is the LL Coulomb Wilson
coefficient for a color singlet heavy quark pair.

For the NLL order approximation it has been frequently stated that there are
no new operators that can contribute and that power counting tells that we
only need to consider ${\cal O}(\alpha_s)$ QCD corrections to the LL 
matching conditions in Eq.~(\ref{Lke}) to account for all electroweak
effects~\cite{Fadin2,Melnikov1}. However, it was also noted 
in~\cite{HoangTeubnerdist} that   
the mismatch between the $t\bar t$ phase space in the full and the effective
theory leads to additional NLL matching corrections for unstable top
quarks. We come back to the role played by these specific NLL contributions
in Sec.\,\ref{sectionrenormalization} and ignore them for the following
considerations.  Thus, concerning NLL effects related to the top
decay only the one-loop QCD corrections to the on-shell top decay width have
to be accounted for~\cite{Jezabek1}. In particular, there are no QCD interference
effects from gluon radiation off the top/antitop quark or its decay
products\,\cite{Fadin2,Melnikov1}. In our approach, which only aims at the
total cross section, one can show that such QCD interference effects do not
even contribute at NNLL order in the 
non-relativistic expansion. 

To discuss the QCD interference
contributions we have to consider ultrasoft gluons,
which carry momenta of order $m_t v^2\sim\Gamma_t$, because they can
interact with a resonant top quark without kicking it off-shell. For the
time-like $A^0$ gluons we already mentioned that their leading interaction with
the quarks can be removed by a field redefinition related to static Wilson
lines~\cite{Korchemsky1,Bauer1}. Moreover, QCD 
gauge invariance ensures that the dominant electroweak matching corrections to the $A^0$
interaction vertex vanish  because we can set the ultrasoft gluon momentum to
zero. Radiative corrections can, however, generate an anomalous interaction in
analogy to the $g-2$ in QED. Yet this interaction is suppressed by a factor
$1/m_t^2$ and cannot contribute to the cross section matrix elements at NNLL
level without even having accounted for additional powers of
the coupling constants. It remains to discuss  the space-like ultrasoft
$\bmA$ gluons 
which couple to the quarks with the $\bmp.\bmA/m_t$ coupling. In pure QCD
space-like ultrasoft gluon exchange contributes to the cross section matrix
elements at NNLL order and also to the renormalization
group running of the $t\bar t$ production operators at the NLL
level~\cite{LMR} through operator mixing.
Accounting for the 
$g^2$-suppression from an additional electroweak loop correction to the
interaction vertex then also leads to a contribution beyond the NNLL order. 

At NNLL order let us first consider whether one needs to account for any
operator in addition to those present already at the LL level. Ultrasoft
gluon interactions have been discussed above. Soft gluon operators first
contribute at the NLL level in pure QCD (for example as corrections to the
Coulomb potential~\cite{amis,amis2}), so $g^2$ electroweak corrections to
these operators are 
beyond NNLL order. Such corrections can also not contribute at NNLL order
through mixing since the Coulomb potential does not cause UV divergences.
It remains to discuss four-quark operators. Because an electroweak loop would 
require a factor $g^4$ in addition to the $1/m_t^2$ suppression for
dimensional reasons, such an operator could not contribute at NNLL order
either~\cite{Beneke2}. 

It remains to discuss $g^2$ corrections to the operators
contributing at the LL level. For the bilinear quark operators in the
effective theory
Lagrangian one obtains the terms shown in Eq.\,(\ref{Lke}). Concerning the
instability of the top quark only the time dilatation correction 
is obtained, and the ${\cal O}(\alpha_s^2)$ and one-loop electroweak
corrections to the on-shell top quark width~\cite{Blokland1,Denner1} have to
be accounted for. The ${\cal O}(\alpha_s^2)$ 
correction to the top width is easy to implement together with the Born and
one-loop QCD results in the width term $\Gamma_t$ and will not be discussed
further in this work. On the other hand, for the Coulomb
potential all dominant $g^2$ 
corrections cancel due to SU(3) gauge invariance because in the first
approximation one can neglect the gluon momentum flowing into the vertex
correction~\cite{Modritsch1,WiseTTbarSusy}. 
The mechanism is equivalent to the gauge cancellation discussed
above for the time-like $A^0$ gluon. The order $g^4$  matching
conditions of the production
operators ${\cal O}_{V,\bbmp}$ and ${\cal O}_{A,\bbmp}$, on the other hand, do
not cancel and have to be determined from matching to the one-loop Standard
Model amplitudes for the process $e^+e^-\to\gamma,Z\to t\bar t$. At this order
the matching computation can be carried out for an on-shell top-antitop
pair at rest. The full set of one-loop electroweak corrections was determined
in Ref.\,\cite{GuthKuehn}. For the examinations in this work we have to
account only for the  $bW^+$ and $\bar b W^-$ cuts because, as shown in 
Sec.\,\ref{sectionrenormalization}, 
only these cuts are relevant for the $bW^+\bar b W^-$ final state that can
interfere with top pair production. Because in Ref.\,\cite{GuthKuehn} only the
sum of 
all contributions was presented, including the $b\bar b$ and $W^+W^-$ cuts, we
rederive the results for the $bW^+$ and $\bar b W^-$ cuts in the next section. 
%Note that these NNLL order corrections related to the instability of the top
%quark cannot be implemented by a simple replacement prescription into
%predictions for stable top quarks.  

%
%
\section{Absorptive Matching Conditions} 
\label{sectionmatchingconditions}

\begin{figure}[t] % figcuts
\begin{center}
\includegraphics[width=14cm]{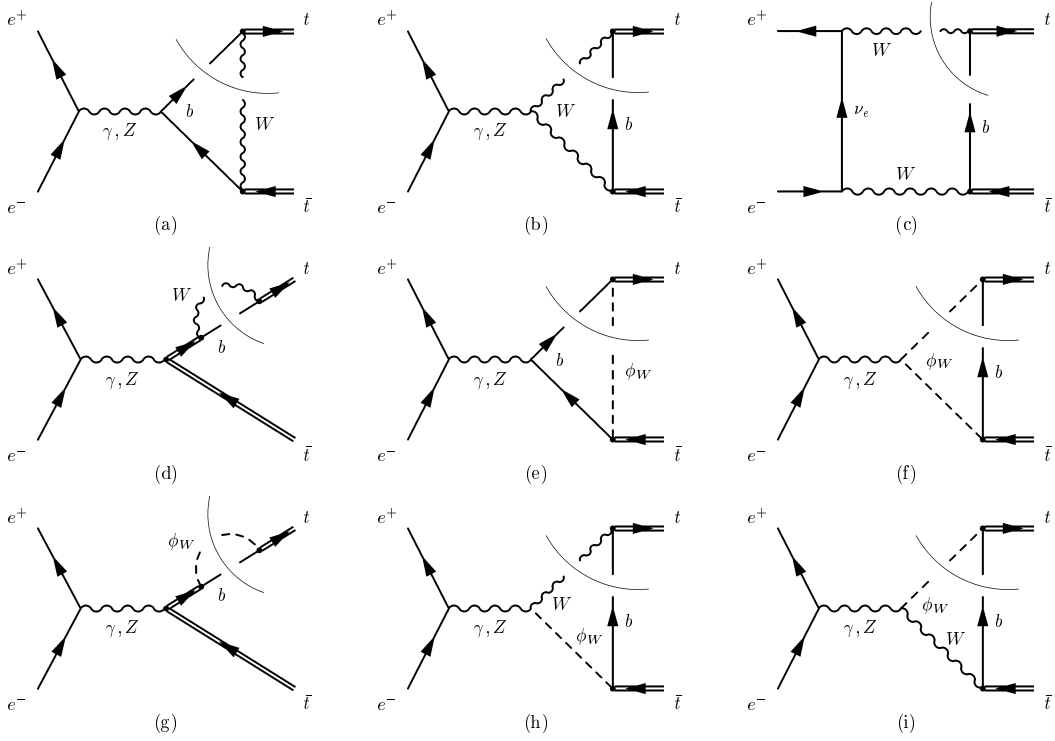}
% \leavevmode
% \epsfxsize=14cm
% \leavevmode
% \epsffile[45 460 545 575]{figures/fig1.ps}
% \vskip  0.0cm
 \caption{
Full theory diagrams in Feynman gauge that have to be considered to determine the 
electroweak absorptive parts in the Wilson coefficients $C_A$ and $C_V$ 
related to the physical $bW^+$ and $\bar b W^-$ intermediate states. Only the
$bW^+$ cut is drawn explicitly. 
 \label{figcuts} }
\end{center}
\end{figure}

The top pair production 
diagrams in the full theory that need to be considered to determine the
absorptive $bW^+$ and $\bar b W^-$ cuts are shown in Fig.\,\ref{figcuts}.
The external (on-shell) top quarks can be taken to be at rest.
%For simplicity only the $bW^+$ cuts are drawn in Fig.\,\ref{figcuts}. 
%
% Fig. figcuts
%
The results for the cuts in the full theory amplitude have the form
\begin{eqnarray}
{\cal A} & = &
i\,\Big[\,\bar v_{e^+}(k^\prime)\,\gamma^\mu
 (i C_V^{\rm bW,abs}+i C_A^{\rm bW,abs}\,\gamma_5)\, u_{e^-}(k)\,\Big]\,
\Big[\,\bar u_t(p)\,\gamma_\mu\,v_{\bar t}(p)\,\Big]
\,,
\label{eett}
\end{eqnarray}
where $k+k^\prime=2p=(2m_t,0)$ and
\begin{eqnarray}
i C_V^{\rm bW,abs} & = & 
-i\,\frac{\alpha^2 \pi |V_{tb}|^2}{12 m_t^2 s_w^2 x(4c_w^2-x)(1+x)}\,
\bigg[\,
\frac{3x(1+x)}{(1-x)}\bigg(1+\frac{x-4}{4s_w^2}\bigg)\,\ln\Big(\frac{2-x}{x}\Big)
\nonumber\\[2mm] & &
+\,Q_e Q_t (1-x)(4-x)(1+2x)(1+x+x^2)
\nonumber\\[3mm] & & 
+\, Q_e(x-1)(1+4x+2x^2+2x^3)
\, +\,  Q_t(1-x)(1+2x)(1+x+x^2)
\nonumber\\[2mm] & &
-\,\frac{1}{2}(1+12x+9x^2+2x^3)
\,+\,\frac{1}{8s_w^2}(2+41x+28x^2-x^3+2x^4)
\,\bigg]
\,,
\nonumber\\[3mm]
i C_A^{\rm bW,abs} & = &
i\,\frac{\alpha^2 \pi |V_{tb}|^2}{12 m_t^2 s_w^2 x(4c_w^2-x)(1+x)}\,
\bigg[\,
\frac{3x(1+x)}{(1-x)}\bigg(1+\frac{x-4}{4s_w^2}\bigg)\,\ln\Big(\frac{2-x}{x}\Big)
\nonumber\\[3mm] & &
+\, Q_t(1-x)(1+2x)(1+x+x^2)
\nonumber\\[2mm] & &
-\, \frac{1}{2}(1+12x+9x^2+2x^3)
\,+\,\frac{1}{8s_w^2}(2+41x+28x^2-x^3+2x^4)
\,\bigg]
\,.
\end{eqnarray}
The results for the charge conjugated process describing top pair
annihilation, on the other hand, read
\begin{eqnarray}
\bar{\cal A} & = &
i\,\Big[\,\bar u_{e^-}(k)\,\gamma^\mu
 (i C_V^{\rm bW,abs}+i C_A^{\rm bW,abs}\,\gamma_5)\, v_{e^+}(k^\prime)\,\Big]\,
\Big[\,\bar v_{\bar t}(p)\,\gamma_\mu\,u_t(p)\,\Big]
\,.
\label{eettadj}
\end{eqnarray}
We used the cutting equations to obtain the result
and checked electroweak gauge invariance by carrying out the computation in
unitary and Feynman gauge. 
In both cases we computed the cut $W$ lines with physical polarizations as well
as with unphysical ones including also the charged Goldstone exchange.
%keeping for the cut W propagator either only
%physical polarizations of unphysical ones which a compensated by an additional
%exchange of charged Goldstone bosons. 
We note that the contributions that
arise from off-shell corrections in the top self-energy graphs are necessary
for electroweak gauge invariance. Since the $b\bar b$ and $W^+W^-$ cuts 
lead to different distinct phase space factors, we found that it is possible
to identify the results also from the formulae given in~\cite{GuthKuehn}. 

It is an important fact that the sign of the imaginary part of the amplitude
does not change in the charge conjugated amplitude. As for the
quark field bilinear terms discussed before in Eq.\,(\ref{Lke}) this is
related to the unitarity of the underlying theory.
It is straightforward to match the amplitudes for the operators ${\cal O}_{V/A,\bbmp}$
and ${\cal O}^\dagger_{V/A,\bbmp}$ to the full theory results in Eqs.\,(\ref{eett})
and (\ref{eettadj}).
The resulting matching conditions at the hard scale read
\begin{eqnarray}
C_V(\nu=1) & = & C_V^{\rm born} + i\,C_V^{\rm bW,abs}
\,,
\nonumber
\\
C_A(\nu=1) & = & C_A^{\rm born} + i\,C_A^{\rm bW,abs}
\,,
\label{Cvaabs}
\end{eqnarray}
where we have included also the Born level contributions from
Eqs.\,(\ref{Cvborn}) and (\ref{Caborn}). We emphasize again that these matching conditions are 
valid for the operators  ${\cal O}_{V/A,\bbmp}$ and ${\cal O}^\dagger_{V/A,\bbmp}$.
In a full treatment of electroweak and QCD effects the coefficients $C_{V/A}$
also include the real parts of the full set of electroweak one-loop diagrams 
indicated in Figs.\,\ref{figcuts} and the QCD matching
corrections known from previous work~\cite{hmst1,c12loop,Hoangc12loopQED}. 
These corrections lead to an energy-independent
multiplicative modification of the cross section normalization which is,
however, not subject of the investigations in this work. The results for the
real parts of the full set of electroweak one-loop diagrams were given
in~\cite{GuthKuehn}.  

\section{Time-Ordered Product and Renormalization} 
\label{sectionrenormalization}

Using the optical theorem the NNLL order corrections to the total cross
section that come from the absorptive one-loop electroweak matching conditions
for the operators ${\cal O}_{V/A,\bbmp}$ and from the time dilatation corrections
can be computed from the 
imaginary part of the $(e^+e^-)(e^+e^-)$ forward scattering amplitude,
\begin{eqnarray}
\sigma_{\rm tot} & \sim &
\frac{1}{s}\, 
\mbox{Im}\left[\,\Big(C_V^2(\nu)+C_A^2(\nu)\Big)\,L^{lk}\,{\cal  A}_1^{lk}\,\right]
\,,
\end{eqnarray}
%time-ordered 
%product of ${\cal O}_{V/A,\bbmp}$ and ${\cal O}_{V/A,\bbmp}^\dagger$,
%\begin{eqnarray}
%\sigma_{\rm tot} & \sim &
%\frac{1}{s}\,\mbox{Im}\,\left[\,
%\tilde{\cal A}_V + \tilde{\cal A}_A
%\,\right]
%\,,
%\end{eqnarray}
where ($k+k^\prime=(\sqrt{s},0)$ and ${\bf \hat e}=\bmk/|\bmk|$) 
\begin{eqnarray}
L^{lk} &=&
       \frac{1}{4}\,\sum\limits_{e^\pm\rm spins}\,
       \Big[\,\bar v_{e^+}(k^\prime)\,\gamma^l\,(\gamma_5)\,u_{e^-}(k)\,\Big]
       \,\Big[\,\bar u_{e^-}(k)\,\gamma^k\,(\gamma_5)\,v_{e^+}(k^\prime)\,\Big]
\nonumber
\\[2mm]
       &=& \frac{1}{2}\,(k+k^\prime)^2\,(\delta^{lk}-\hat e^l  \hat e^k)
\end{eqnarray}
is the spin-averaged lepton tensor and ($\hat{q}\equiv(\sqrt{s}-2m_t,0)$)
%\begin{eqnarray}
%k^\mu=(\frac{\sqrt{s}}{2},\frac{\sqrt{s}}{2}\,{\mathbf\hat{e}})\\
%{k^\prime}^\mu=(\frac{\sqrt{s}}{2},-\frac{\sqrt{s}}{2}\,{\mathbf\hat{e}})
%\end{eqnarray}
\begin{eqnarray}
{\cal A}_1^{lk} &=& 
i\, \sum\limits_{\mbox{\scriptsize\boldmath $p$},\mbox{\scriptsize\boldmath $p'$}} \int\! d^4x\:
e^{-i\hat q \cdot x}\:
\Big\langle\, 0\,\Big|\,T\,
{{\cal O}_{\mbox{\scriptsize\boldmath $p$},1}^l}^{\!\!\!\dagger} (0)\, 
{\cal O}_{\mbox{\scriptsize\boldmath $p'$},1}^k (x)\Big|\,0\,\Big\rangle
\nonumber\\[2mm] & = &
2\,N_c\,\delta^{lk}\,G^0(a,v,m_t,\nu)
\end{eqnarray}
is the time-ordered product of the $t\bar t$ production and
annihilation operators ${\cal O}_{\bbmp,1}^j$ and ${{\cal
    O}_{\bbmp,1}^j}^{\!\!\!\dagger}$~\cite{hmst1}.
%where  ($\hat{q}\equiv(\sqrt{s}-2m,0)$)
%\begin{eqnarray}
%\tilde{\cal A}_{j} & = &
%C^2_{j}(\nu)\,\,\frac{1}{4}\sum_{e^\pm {\rm spins}} \,\,\,
%\sum\limits_{\mbox{\scriptsize\boldmath $p$},\mbox{\scriptsize\boldmath $p'$}}
%i \int\! d^4x\: e^{-i \hat{q} \cdot x}\:
% \Big\langle\,0\,\Big|\, T\, {\cal O}_{j,\bbmp}(x)\,{\cal O}^\dagger_{j,\bbmp^\prime}(0)
% \Big|\,0\,\Big\rangle \,.
%\end{eqnarray}
In dimensional regularization the result reads
\begin{eqnarray}
\Delta\sigma_{\rm tot}^{\Gamma,1} & = &
2\,N_c\,\mbox{Im}\bigg\{\,
2i\,\big[\,C_V^{\rm born}\,C_V^{\rm bW,abs}+C_A^{\rm
  born}\,C_A^{\rm bW,abs}\,\big]\,G^0(a,v,m_t,\nu)
\nonumber\\[2mm] & &
+\,\big[\,(C_V^{\rm born})^2+(C_A^{\rm born})^2\,\big]\,
\delta G^0_\Gamma(a,v,m_t,\nu)
\,\bigg\}
\,,
\label{dsignaNNLL}
\end{eqnarray}
where $a\equiv -{\cal V}_c^{(s)}(\nu)/4\pi=C_F\alpha_s(m_t\nu)$. The term $G^0$ is
the zero-distance 
S-wave Green function of the non-relativistic Schr\"odinger
equation which is obtained from the LL order terms in the Lagrangian 
shown in Eqs.\,(\ref{Lke}) and (\ref{Lpot}).
 In dimensional regularization it has
the form~\cite{hmst1} 
\begin{eqnarray}
 G^0(a,v,m_t,\nu) & = &
 \frac{m_t^2}{4\pi}\left\{\,
 i\,v - a\left[\,\ln\left(\frac{-i\,v}{\nu}\right)
 -\frac{1}{2}+\ln 2+\gamma_E+\psi\left(1\!-\!\frac{i\,a}{2\,v}\right)\,\right]
 \,\right\}
 \nonumber \\ & &
 +\,\frac{m_t^2\,a}{4 \pi}\,\,\frac{1}{4\,\epsilon}
\,,
% -\frac{1}{6}\,\right) 
\label{deltaGCoul}
\end{eqnarray}  
where
\begin{eqnarray}
v & = &
\sqrt{\frac{\sqrt{s}-2(m_t+\delta m_t)+i\Gamma_t}{m_t}}
\,,
\end{eqnarray}
$\sqrt{s}$ being the c.\,m.\,\,energy.
The term $\delta G^0_\Gamma$ represents the corrections originating
from the time dilatation correction in Eq.\,(\ref{Lke}) and reads
\begin{eqnarray}
\delta G^0_\Gamma(a,v,m_t,\nu) & = &
-i\,\frac{\Gamma_t}{2m_t}\,\bigg[\,
1+\frac{v}{2}\frac{\partial}{\partial v} + a\frac{\partial}{\partial a}
\,\bigg]\,G^0(a,v,m_t,\nu)
\,.
\label{deltaGGamma}
\end{eqnarray}
Note that the Wilson coefficients $C_{V/A}$ do not run at LL order,
so only the matching conditions at $\nu=1$ appear in Eq.\,(\ref{dsignaNNLL}).

\begin{figure}[t] % figinterfer
\begin{center}
\includegraphics[width=14cm]{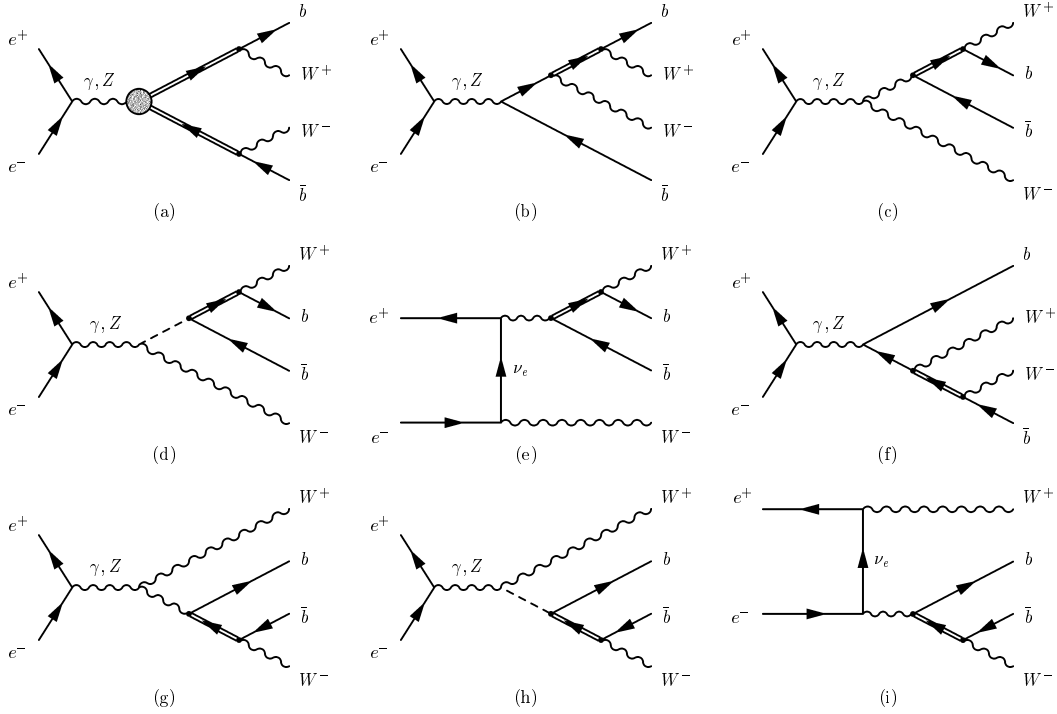}
% \leavevmode
% \epsfxsize=14cm
% \leavevmode
% \epsffile[45 460 545 575]{figures/fig2.ps}
% \vskip  0.0cm
 \caption{
Full theory Feynman diagrams describing the process 
$e^+e^-\to bW^+\bar b W^-$ with one or two intermediate top or antitop quark
propagators. 
%The top/antitop lines are indicated with double lines, the bottom
%and W lines are drawn as straight single and wavy lines, respectively.
The circle in diagram (a) represents the QCD form
factors for the $t\bar t$ vector and axial-vector currents.
 \label{figinterfer} }
\end{center}
\end{figure}

It is straightforward to check that the terms proportional to
$C_{V/A}^{\rm bW,abs}$ in Eq.\,(\ref{dsignaNNLL}) are in agreement with the
full theory matrix elements from the interference
between the double-resonant amplitudes for the process
$e^+e^-\to t\bar t\to bW^+\bar b W^-$ (Fig.\,\ref{figinterfer}a) and the single-resonant amplitudes describing the processes $e^+e^-\to t+\bar b W^-\to
bW^+\bar b W^- $  and 
$e^+e^-\to bW^+\bar t\to bW^+\bar b W^-$ (Figs.\,\ref{figinterfer}b-i) in the
$t\bar t$ threshold limit for $m_t\to\infty$. 
%
% Fig. figinterfer
%
Note that diagram (a) dominates in the non-relativistic limit due to two
resonant top/antitop lines, while diagrams (b-i) are $v^2$-suppressed having
only one resonant top/antitop line. Diagram (a) also contains a
subleading $v^2$-suppressed contribution that has to be accounted for.
Diagrams with no top/antitop line are
suppressed by $v^4$ and do not need to be considered. This also means that
pure background diagrams containing no intermediate top quark can be
neglected at this order. We also note that to find literal
agreement between full and effective theory matrix elements one has to replace
the $i\epsilon$ terms in the resonant full theory top 
propagators by the Breit-Wigner term $i m_t\Gamma_t/2$. The circle
shown in Fig.\,\ref{figinterfer}a represents the QCD form factors for
the $t \bar t$ vector and axial-vector currents. In the
non-relativistic limit they reduce to the insertions of Coulomb
potentials described by the higher order terms in
Eq.\,(\ref{deltaGCoul}). Due to the cancellation of the QCD
interference effects caused by gluons with ultrasoft momenta there are
no further QCD corrections in the non-relativistic limit.

An interesting new conceptual aspect of the corrections shown in
Eq.\,(\ref{dsignaNNLL}) is that they have UV $1/\epsilon$-divergences that
arise from a logarithmic high energy behavior of the top-antitop effective
theory phase space integration for matrix elements containing a single
insertion of the Coulomb potential. In the forward scattering amplitude these 
UV divergences arise because the imaginary parts of the matching conditions
of Eqs.\,(\ref{Cvaabs}) lead to a dependence on the real part of $G^0$  
(see Eq.~(\ref{deltaGCoul})). In the full theory this logarithmic behavior is
regularized by the top quark mass. While phase space logarithms are known in
the literature and can be resummed with renormalization group
techniques~\cite{Bauer2}, 
the divergences here are specific since they would not exist if the top quark
were approximated as being stable. In particular, the UV divergences
from the time dilatation 
corrections arise from the Breit-Wigner-type high energy behavior of the
effective theory top propagator in Eq.\,(\ref{toppropagator}) which differs
from the one for a 
stable particle. Likewise, the interference effects described by the absorptive
electroweak matching conditions for the operators ${\cal O}_{V,\bbmp}$ and 
${\cal O}_{A,\bbmp}$ would not have to be taken into account if the top quarks
were stable particles. UV divergences of the same kind have already been
observed and described before in the NNLL relativistic corrections to
the S-wave zero-distance Green function if the unstable propagator in
Eq.\,(\ref{toppropagator}) is
used~\cite{hmst1,HoangTeubnerdist,HoangTeubner}. 
%\footnote{We recall
%that applying the prescription $E\to E+i\Gamma_t$ to predictions
%for a stable non-relativistic top quark reproduce the computation where the
%propagator $i/(p^0 - {\itbf{p}}^2/(2m) + i\Gamma_t/2)$ is used from the
%beginning.}  
For the P-wave zero-distance Green function, which is generated by $t\bar t$
production through an axial-vector current from the Z-exchange and that
contributes only at NNLL order,
a similar UV divergence arises already at leading order in the non-relativistic
expansion. Like for the case of the time dilatation corrections these
divergences originate from the modified high energy behavior of the unstable top
propagator. 
We believe it is evident that these divergences do not represent a
deficiency of the effective theory, because the concept of separating resonant
and non-resonant fluctuations appears to be the only practical way to make
systematic predictions involving unstable particles. Thus these 
UV divergences should be handled with the  renormalization techniques
known from effective theories for stable particles. The only difference is
that the renormalization procedure will involve operators having non-hermitian
Wilson coefficients. 

The operators that are renormalized by the UV divergences displayed in
Eq.\,(\ref{dsignaNNLL}) are the two $(e^+e^-)(e^+e^-)$ operators 
\begin{eqnarray}
\tilde{\cal O}_V & = &
-\,\big[\,\bar e\,\gamma^\mu\,e\,\big]\,
\big[\,\bar e\,\gamma_\mu\,e\,\big]
\,,
\\[2mm]
\tilde{\cal O}_A & = &
-\,\big[\,\bar e\,\gamma^\mu\,\gamma_5 \, e\,\big]\,
\big[\,\bar e\,\gamma_\mu\,\gamma_5\,e\,\big]
\,,
\end{eqnarray}
which give  the additional contribution 
$\tilde\Delta {\cal L} = \tilde C_V \tilde{\cal O}_V + \tilde C_A
\tilde{\cal O}_A$
to the effective theory Lagrangian,  $\tilde C_{V/A}$ being the Wilson
coefficients.
Because in this work we neglect QED effects, the electron and
positron act as classic fields and therefore  $\tilde C_{V}$ and $\tilde
C_{A}$ run only through mixing due to UV divergences such as in 
Eq.\,(\ref{dsignaNNLL}). 
Since only the imaginary parts of the coefficients $\tilde
C_{V/A}$ can contribute to the total cross section through the optical theorem
we neglect the real contributions in the following.
%Since the operators $\tilde{\cal O}_{V/A}$ are renormalized by the divergences
%in the time-ordered product of Eq.\,(\ref{dsignaNNLL}) they are, as expected,
%also non-hermitian. 
Using the standard $\msb$ subtraction procedure the (non-hermitian)
counterterms of the renormalized $\tilde{\cal O}_{V/A}$ 
operators read
\begin{eqnarray}
\delta \tilde C_V & = &
i\,\frac{N_c m_t^2}{32 \pi^2 \epsilon}\,
\bigg[ (C_V^{\rm born})^2 \frac{\Gamma_t}{m_t} 
  +2 C_V^{\rm born} C_V^{\rm bW,abs}  
\bigg]\,
{\cal V}_c^{(s)}(\nu) 
\nonumber \\[2mm] & & 
+\,i\,\frac{N_c m_t^2}{32 \pi^2 \epsilon}\,
(C_V^{\rm born})^2 \frac{\Gamma_t}{m_t} \,
\Big[ \Big( 2c_2(\nu) -1\Big){\cal V}_c^{(s)}(\nu) + {\cal V}_r^{(s)}(\nu)  \Big]
\nonumber \\[2mm] & & 
+\,i\,\frac{N_c m_t^2}{48 \pi^2 \epsilon}\,
(C_V^{\rm ax})^2 \frac{\Gamma_t}{m_t} \,{\cal V}_c^{(s)}(\nu)
\,,
\nonumber\\[4mm]
\delta \tilde C_A & = & 
i\,\frac{N_c m_t^2}{32 \pi^2 \epsilon}\,
\bigg[ (C_A^{\rm born})^2 \frac{\Gamma_t}{m_t} 
  +2 C_A^{\rm born} C_A^{\rm bW,abs}  
\bigg]\,
{\cal V}_c^{(s)}(\nu) 
\nonumber \\[2mm]& & 
+\,i\,\frac{N_c m_t^2}{32 \pi^2 \epsilon}\,
(C_A^{\rm born})^2 \frac{\Gamma_t}{m_t} \,
\Big[ \Big( 2c_2(\nu) -1\Big){\cal V}_c^{(s)}(\nu) + {\cal V}_r^{(s)}(\nu)  \Big]
\nonumber \\[2mm] & & 
+\,i\,\frac{N_c m_t^2}{48 \pi^2 \epsilon}\,
(C_A^{\rm ax})^2 \frac{\Gamma_t}{m_t} \,{\cal V}_c^{(s)}(\nu)
\,,
\label{Otildecounter}
\end{eqnarray}
where the respective first term on the RHS's
subtract the $1/\epsilon$ divergences shown in
Eq.\,(\ref{dsignaNNLL}) and the other terms account for the UV divergences in the
P-wave Green function and the NNLL corrections of the S-wave Green function
computed in Ref.~\cite{hmst1}. Here, $C_{V/A}^{\rm ax}$ are Born level Wilson
coefficients of operators describing top pair production in a P-wave
(originating from pure Z exchange), ${\cal V}_r^{(s)}$ is the color singlet
coefficient of the potential
$(\bmp^2+\bmp^{\prime\,2})/(2m_t(\bmp-\bmp^\prime)^2)$~\cite{HoangStewartultra}
and $c_2$ the coefficient of the $\bmp^2$-suppressed S-wave production 
current~\cite{hmst1}. The explicit formulae read
\begin{eqnarray}
 C_{V}^{\rm ax} & = & 
\frac{\alpha\pi}{m_t^2(4c_w^2-x)}\,\bigg[\, Q_e+\frac{1}{4s_w^2}  \,\bigg]
\,,
\nonumber\\[2mm]
 C_{A}^{\rm ax} & = &
-\frac{\alpha\pi}{4 s_w^2 m_t^2(4c_w^2-x)}
\,,
\nonumber\\[2mm]
{\cal V}_r^{(s)}(\nu) & = & -4\pi C_F \alpha_s(m_t) z\,
\left[ 1+\frac{8C_A}{3\beta_0}\ln(2-z)
\right]
\,,
\nonumber\\[2mm]
c_2(\nu) & = & -\frac{1}{6} 
- \frac{8
  C_F}{3\beta_0}\ln\Big(\frac{z}{2-z}\Big)
\,,
\nonumber\\[2mm]
z & \equiv & \frac{\alpha_s(m_t\nu)}{\alpha_s(m_t)}
\,.
\end{eqnarray}
The resulting renormalization group equations for the Wilson coefficients
$\tilde C_{V/A}$ have the form
\begin{eqnarray}
\frac{d \tilde C_V(\nu)}{d\ln\nu} & = &
i\,\frac{N_c m_t^2}{8\pi^2}\,\bigg\{\,
(C_V^{\rm born})^2\,\frac{\Gamma_t}{m}\,\Big( 
 2 c_2(\nu) {\cal V}^{(s)}_c(\nu) + {\cal V}_r^{(s)}(\nu)\Big) 
+ 2 C_V^{\rm born} C_V^{\rm bW,abs}  {\cal V}^{(s)}_c(\nu)
\,\bigg\}
\nonumber\\[2mm] & &
+\,i\,\frac{N_c m_t^2}{12\pi^2}\,\bigg\{\,
(C_V^{\rm ax})^2\,\frac{\Gamma_t}{m_t}\,{\cal V}^{(s)}_c(\nu)
\,\bigg\}
\,,
\nonumber\\[4mm]
\frac{d \tilde C_A(\nu)}{d\ln\nu} & = &
i\,\frac{N_c m_t^2}{8\pi^2}\,\bigg\{\,
(C_A^{\rm born})^2\,\frac{\Gamma_t}{m}\,\Big( 
 2 c_2(\nu) {\cal V}^{(s)}_c(\nu) + {\cal V}_r^{(s)}(\nu)\Big) 
+ 2 C_A^{\rm born} C_A^{\rm bW,abs}  {\cal V}^{(s)}_c(\nu)
\,\bigg\}
\nonumber\\[2mm] & &
+\,i\,\frac{N_c m_t^2}{12\pi^2}\,\bigg\{\,
(C_A^{\rm ax})^2\,\frac{\Gamma_t}{m_t}\,{\cal V}^{(s)}_c(\nu)
\,\bigg\}
\,,
\end{eqnarray}
and the solutions for scales below $m_t$ ($\nu<1$) read
\begin{eqnarray}
\tilde C_V(\nu)& = & \tilde C_V(1) + 
i\,\frac{2 N_c m_t^2 C_F}{3\beta_0}\,\bigg\{
\bigg[ \Big( (C_V^{\rm born})^2+ (C_V^{\rm ax})^2 \Big)\,\frac{\Gamma_t}{m_t}
 + 3 C_V^{\rm born} C_V^{\rm bW,abs}\bigg]\,\ln(z)
\nonumber\\[2mm] & &\qquad
 - \frac{4C_F}{\beta_0}\,\frac{\Gamma_t}{m_t}\,(C_V^{\rm born})^2\ln^2(z)
 + \frac{4(C_A+2C_F)}{\beta_0}\,\frac{\Gamma_t}{m_t} \,(C_V^{\rm born})^2\rho(z)
\bigg\}
\,,
\nonumber\\[4mm] 
\tilde C_A(\nu)& = & \tilde C_A(1) + 
i\,\frac{2 N_c m_t^2 C_F}{3\beta_0}\,\bigg\{
\bigg[ \Big( (C_A^{\rm born})^2+ (C_A^{\rm ax})^2 \Big)\,\frac{\Gamma_t}{m_t}
 + 3 C_A^{\rm born} C_A^{\rm bW,abs}\bigg]\,\ln(z)
\nonumber\\[2mm] & &\qquad
 - \frac{4C_F}{\beta_0}\,\frac{\Gamma_t}{m_t}\,(C_A^{\rm born})^2\ln^2(z)
 + \frac{4(C_A+2C_F)}{\beta_0}\,\frac{\Gamma_t}{m_t} \,(C_A^{\rm born})^2\rho(z)
\bigg\}
\,,
\label{tildeCav}
\end{eqnarray}
where
\begin{eqnarray}
\rho(z) & = & \frac{\pi^2}{12}-\frac{1}{2}\ln^22 + \ln2 \ln(z) - {\rm
  Li}_2\left(\frac{z}{2}\right)
%\,,
%\nonumber\\ 
%z & \equiv & \frac{\alpha_s(m_t \nu)}{\alpha_s(m_t)}
\,,
\end{eqnarray}
and the $\tilde C_{V/A}(1)$ are the hard matching conditions, which
are
presently unknown.
Finally, the contribution of the operators $\tilde {\cal O}_{V/A}$ to the
total cross section reads
\begin{eqnarray}
\Delta\sigma_{\rm tot}^{\Gamma,2} & = & 
\mbox{Im}\Big[\,{\tilde C_V} +{\tilde C_A}\,\Big]
\,.
\end{eqnarray}
Parametrically $\Delta\sigma_{\rm tot}^{\Gamma,2}$ is of order
$g^6$. Comparing this with the LL cross section which counts as $g^4 v\sim
g^5$ we see that it constitutes a NLL contribution. This is also evident from
the fact that the corresponding UV divergences were generated in NNLL order
effective theory matrix
elements. The correction $\Delta\sigma_{\rm tot}^{\Gamma,2}$ is
energy-independent, but it is scale-dependent and
compensates the logarithmic scale-dependence in the NNLL order matrix
elements. 
%In this respect the matching conditions $\tilde C_{V/A}(\nu=1)$ that
%appear in Eqs.\,(\ref{tildeCav}) are less significant because they can be
%suppressed if the line-shape measurements are carried out relative to the
%$bW^+\bar b W^-$ cross section at an energy well below the peak position (but
%still in the non-relativistic regime). 
As mentioned before, the matching conditions $\tilde C_{V/A}(\nu=1)$ are presently
unknown and we therefore set them to zero
in the numerical analysis presented below. We note, however, that it was shown
in~\cite{HoangTeubnerdist} that the difference between the full theory phase
space (which is cut off by the large, but finite $m_t$) and the effective theory
phase space (which is infinite in the computation of the forward scattering
amplitude) contributes to $\tilde C_{V/A}(\nu=1)$ and also represents a NLL
effect.

\section{Numerical Analysis}
\label{sectionanalysis}
%
%

%
% Fig. figanalysis
%
\begin{figure}[t] % figanalysis
\begin{center}
 \leavevmode
 \epsfxsize=10cm
 \leavevmode
 \epsffile[100 430 580 730]{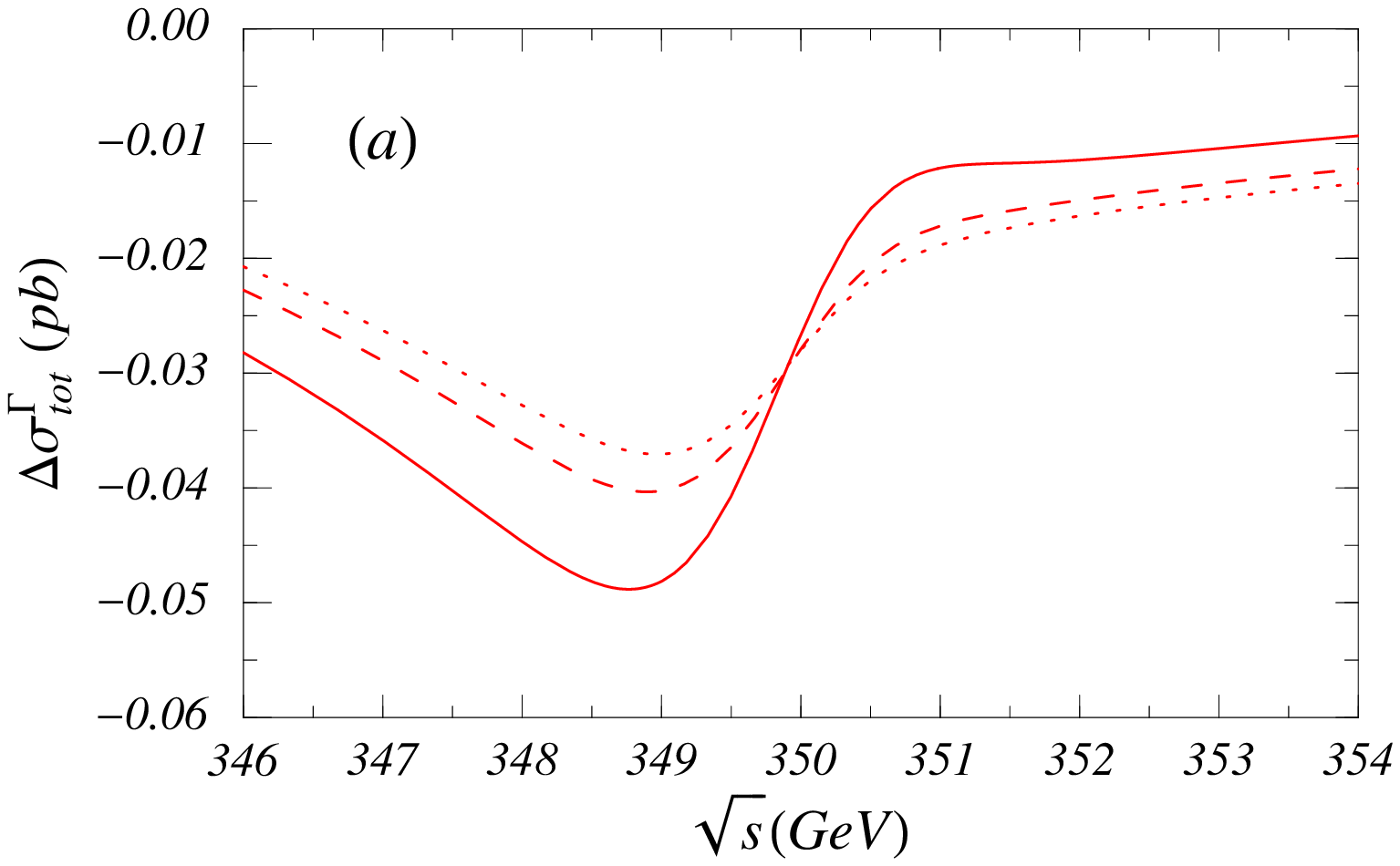}\\
 \epsfxsize=10cm
 \leavevmode
 \epsffile[100 430 580 730]{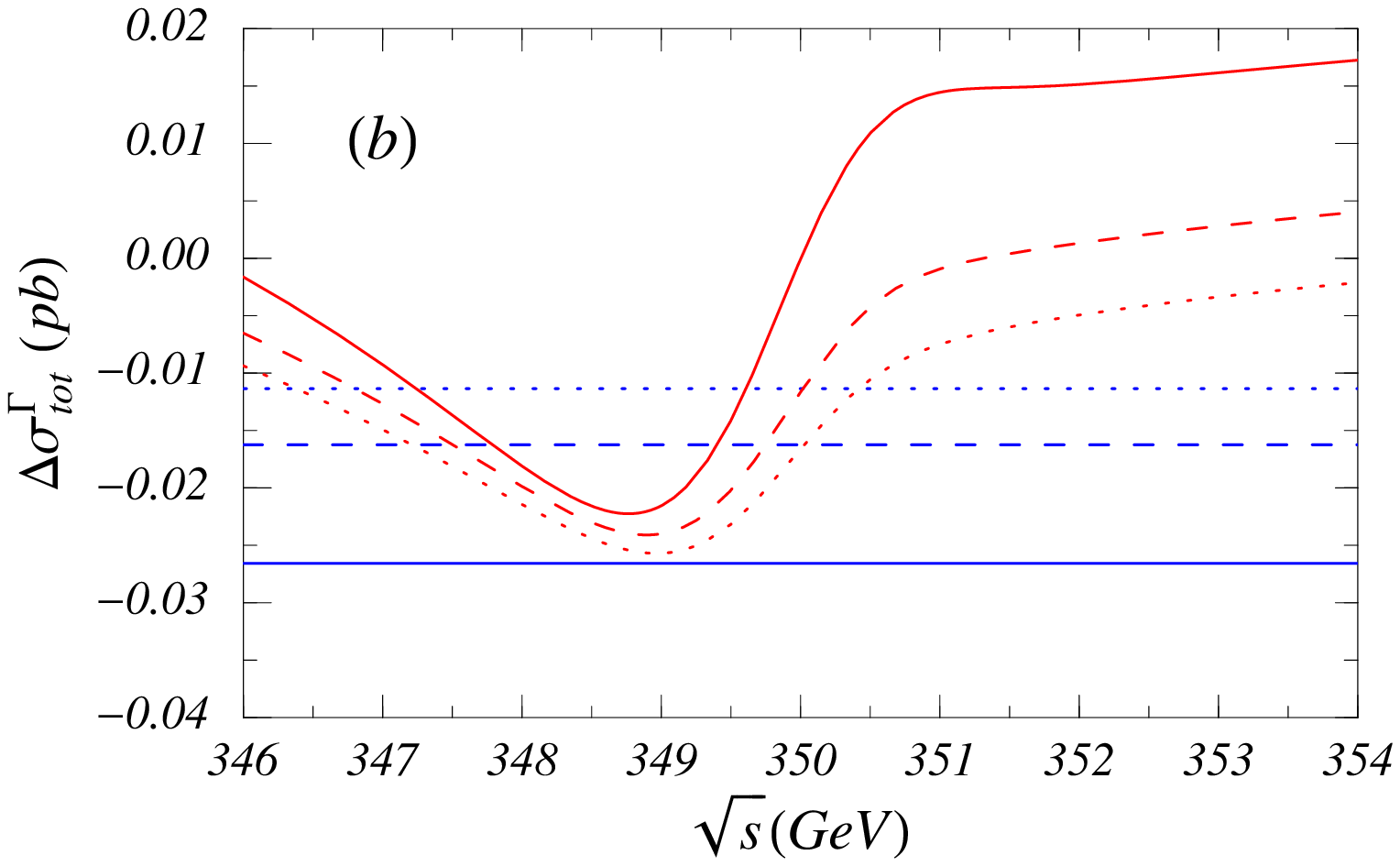}\\
 \vskip  0.0cm
 \caption{
The corrections $\Delta\sigma_{\rm tot}^{\Gamma,1}$  and 
$\Delta\sigma_{\rm tot}^{\Gamma,2}$ in $pb$ for $M_{\rm 1S}=175$~GeV,
$\alpha=1/125.7$, $s_w^2=0.23120$, $V_{tb}=1 $, $M_W=80.425$~GeV,
$\Gamma_t=1.43$~GeV and $\nu=0.1$ (solid
curves),  $0.2$ (dashed curves) and $0.3$ (dotted curves) in the energy
range $346~\mbox{GeV} < \sqrt{s} < 354$~GeV.
Panel (a) shows the sum of both corrections and panel (b) the individual size
of $\Delta\sigma_{\rm tot}^{\Gamma,1}$ (energy-dependent lines) and 
$\Delta\sigma_{\rm tot}^{\Gamma,2}$ (straight lines).
 \label{figanalysis} }
\end{center}
\end{figure}

In Fig.\,\ref{figanalysis} we have plotted 
$\Delta\sigma_{\rm tot}^{\Gamma,1}$  and $\Delta\sigma_{\rm tot}^{\Gamma,2}$ 
in picobarn in the 1S mass scheme~\cite{HoangTeubnerdist,Hoangupsilon} for
$M_{\rm 1S}=175$~GeV, $\alpha=1/125.7$, $s_w^2=0.23120$, $V_{tb}=1$ and
$M_W=80.425$~GeV with the renormalization scaling parameter $\nu=0.1$ (solid
curves),  $0.2$ (dashed curves) and $0.3$ (dotted curves). 
The divergences in $\Delta\sigma_{\rm tot}^{\Gamma,1}$ are
subtracted minimally. For the QCD
coupling we used $\alpha_s(M_Z)=0.118$ as an input and employed 4-loop
renormalization group running. Note that in the 1S scheme 
$\delta m_t=M_{\rm 1S}({\cal V}_c^{(s)}(\nu)/4\pi)^2/8$. In
Fig.\,\ref{figanalysis}a the sum of 
$\Delta\sigma_{\rm tot}^{\Gamma,1}$  and $\Delta\sigma_{\rm tot}^{\Gamma,2}$
is shown while in Fig.\,\ref{figanalysis}b both contributions are presented
separately. For the top 
quark width we adopted the value $\Gamma_t=1.43$~GeV.\footnote{
In a  complete analysis of electroweak effects the top quark
width depends on the input parameters given above and is not an independent
parameter. For the purpose of the numerical analysis in this work, however,
our treatment is sufficient.}
We find that the sum of the corrections is negative and shows a moderate $\nu$
dependence. 
Compared to the most recent NNLL QCD predictions for the total cross
section~\cite{HoangEpiphany} the corrections are around $-10\%$
for energies below the peak, between $-2\%$ and $-4\%$ close to the peak and
about $-2\%$ above the peak. Their magnitude is
comparable to the NNLL QCD corrections. Interestingly, they partly compensate  
the sizeable positive QCD corrections found
in~\cite{Hoang3loop,HoangEpiphany}. The peculiar energy dependence of the
corrections, caused by the dependence on the real part of the Green function
$G^0$, also leads to a slight displacement of the peak position. Relative to
the peak position of the LL cross section one obtains a shift of $(30,35,47)$~MeV
for $\nu=(0.1,0.2,0.3)$. This shift is comparable to the expected
experimental uncertainties of the top mass measurements from the
threshold scan~\cite{TTbarsim}.

\section{Conclusion}
\label{sectionconclusion}

We have determined electroweak corrections to the NNLL top pair threshold
cross section in $e^+e^-$ annihilation related to the instability of the
top quark. Our approach is closely related to the
treatment of absorptive processes in the optical theory. It includes the
effects of the top instability by accounting for the absorptive parts in
electroweak corrections to the effective theory matching conditions
that are related to  
the observable $bW^+\bar b W^-$ final state. The matching conditions render
the NRQCD Lagrangian non-hermitian, but they allow for the determination of
the total cross section using the $e^+e^-\to e^+e^-$ forward scattering
matrix element and the optical theorem. We have shown that the absorptive
parts of the electroweak matching conditions for the $t\bar t$ production and
annihilation operators describe the interference of the double-resonant
amplitude for $e^+e^-\to t\bar t\to bW^+\bar b W^-$ with the single-resonant
(and $v^2$-suppressed) amplitudes for  
$e^+e^-\to b W^+\bar t\to bW^+\bar b W^-$ and  
$e^+e^-\to t\bar b W^-\to bW^+\bar b W^-$. We have also shown that at NNLL
order there are no further interference effects caused by the exchange and
radiation of ultrasoft gluons. 

The novel feature of the NNLL corrections is that they lead to new
UV divergences. These divergences originate from the logarithmic high energy
behavior of the $t\bar t$ phase space in the effective theory forward
scattering amplitudes, that is caused by the interferences and by the modified
propagators of an unstable top quark. The divergences renormalize
$(e^+e^-)(e^+e^-)$ operators that contribute to the forward scattering
amplitude already at NLL order. The corrections determined in this work
slightly modify the cross section shape and are comparable to the known NNLL
QCD corrections. The size of the corrections shows that a complete
treatment of all NNLL electroweak effects is desirable.

\begin{acknowledgments} 
A.H. thanks the Aspen Center for
Physics where this work was initiated and I.\,Stewart for collaboration 
during the initial stages of the project. We thank A.~Manohar for comments to
the manuscript.
\end{acknowledgments}

%\newpage
\appendix

%Bibliography


\begin{thebibliography}{}


\bibitem{Kuehn1}
I.~I.~Bigi, Y.~L.~Dokshitzer, V.~Khoze, J.~K\"uhn and P.~Zerwas,
%``Production And Decay Properties Of Ultraheavy Quarks,''
Phys.\ Lett.\ B {\bf 181}, 157 (1986).
%%CITATION = PHLTA,B181,157;%%


\bibitem{Fadin1}
V.~S.~Fadin and V.~A.~Khoze,
%``Threshold Behavior Of Heavy Top Production In E E- Collisions,''
JETP Lett.\  {\bf 46}, 525 (1987).
%%CITATION = JTPLA,46,525;%%


\bibitem{Habilitation}
A.~H.~Hoang,
%``Heavy quarkonium dynamics,''
arXiv:hep-ph/0204299.
%%CITATION = HEP-PH 0204299;%%


\bibitem{CKMworkshop}
M.~Battaglia {\it et al.},
%``The CKM matrix and the unitarity triangle,''
arXiv:hep-ph/0304132.
%%CITATION = HEP-PH 0304132;%%



\bibitem{TTbarsim}
M.~Martinez and R.~Miquel,
%``Multi-parameter fits to the t anti-t threshold observables at a future  e+
%e- linear collider,'' 
Eur.\ Phys.\ J.\ C {\bf 27}, 49 (2003)
[arXiv:hep-ph/0207315].
%%CITATION = HEP-PH 0207315;%%


\bibitem{synopsis}
A.~H.~Hoang {\it et al.},
%``Top-antitop pair production close to threshold: Synopsis of recent
%NNLO  results,'' 
in Eur.\ Phys.\ J.\ direct C {\bf 2}, 1 (2000)
[arXiv:hep-ph/0001286].
%%CITATION = HEP-PH 0001286;%%


\bibitem{Cinabro1}
D.~Cinabro,
%``The t anti-t threshold and machine parameters at the NLC,''
arXiv:hep-ex/0005015.
%%CITATION = HEP-EX 0005015;%%


\bibitem{LMR} 
M.~Luke, A.~Manohar and I.~Rothstein,
%``Renormalization group scaling in nonrelativistic QCD,''
Phys.\ Rev.\ D {\bf 61}, 074025 (2000)
[arXiv:hep-ph/9910209].
%%CITATION = HEP-PH 9910209;%%


\bibitem{HoangStewartultra}
A.~H.~Hoang and I.~W.~Stewart,
%``Ultrasoft renormalization in non-relativistic QCD. ((V)) ((W)),''
Phys.\ Rev.\ D {\bf 67}, 114020 (2003)
[arXiv:hep-ph/0209340].
%%CITATION = HEP-PH 0209340;%%


\bibitem{Manohar1}
A.~V.~Manohar and I.~W.~Stewart,
%``Running of the heavy quark production current and 1/v potential in QCD,''
Phys.\ Rev.\ D {\bf 63}, 054004 (2001)
[arXiv:hep-ph/0003107].
%%CITATION = HEP-PH 0003107;%%


\bibitem{Pineda1}
A.~Pineda,
%``Next-to-leading-log renormalization-group running in heavy-quarkonium
%creation and annihilation,''
Phys.\ Rev.\ D {\bf 66}, 054022 (2002)
[arXiv:hep-ph/0110216].
%%CITATION = HEP-PH 0110216;%%


\bibitem{Pineda2}
A.~Pineda,
%``Renormalization group improvement of the NRQCD Lagrangian and heavy
%quarkonium spectrum,''
Phys.\ Rev.\ D {\bf 65}, 074007 (2002)
[arXiv:hep-ph/0109117].
%%CITATION = HEP-PH 0109117;%%


\bibitem{Hoang3loop}
A.~H.~Hoang,
%``Three-loop anomalous dimension of the heavy quark pair production  current
%in non-relativistic QCD,''
Phys.\ Rev.\ D {\bf 69}, 034009 (2004)
[arXiv:hep-ph/0307376].
%%CITATION = HEP-PH 0307376;%%


\bibitem{Peninc1}
A.~A.~Penin, A.~Pineda, V.~A.~Smirnov and M.~Steinhauser,
%``Spin dependence of heavy quarkonium production and annihilation rates:
%Complete next-to-next-to-leading logarithmic result,''
Nucl.\ Phys.\  B {\bf 699}, 183 (2004)
[arXiv:hep-ph/0406175].
%%CITATION = HEP-PH 0406175;%%


\bibitem{hmst}
A.~H.~Hoang, A.~V.~Manohar, I.~W.~Stewart and T.~Teubner,
%``A renormalization group improved calculation of top quark
%production  near threshold,'' 
Phys.\ Rev.\ Lett.\  {\bf 86}, 1951 (2001)
[arXiv:hep-ph/0011254].
%%CITATION = HEP-PH 0011254;%%


\bibitem{hmst1}
A.~H.~Hoang, A.~V.~Manohar, I.~W.~Stewart and T.~Teubner,
%``The threshold t anti-t cross section at NNLL order,''
Phys.\ Rev.\ D {\bf 65}, 014014 (2002)
[arXiv:hep-ph/0107144].
%%CITATION = HEP-PH 0107144;%%


\bibitem{HoangEpiphany}
A.~H.~Hoang,
%``Top Pair Production at Threshold and Effective Theories,''
Acta Phys.\ Polon.\ B {\bf 34}, 4491 (2003)
[arXiv:hep-ph/0310301].
%%CITATION = HEP-PH 0310301;%%


\bibitem{GuthKuehn}
R.~J.~Guth and J.~H.~K\"uhn,
%``Top quark threshold and radiative corrections,''
Nucl.\ Phys.\ B {\bf 368}, 38 (1992).
%%CITATION = NUPHA,B368,38;%%


\bibitem{HoangTeubnerdist}
A.~H.~Hoang and T.~Teubner,
%``Top quark pair production close to threshold: Top mass, width and  momentum
%distribution,''
Phys.\ Rev.\ D {\bf 60}, 114027 (1999)
[arXiv:hep-ph/9904468].
%%CITATION = HEP-PH 9904468;%%

\bibitem{Beneke1}
M.~Beneke, A.~P.~Chapovsky, A.~Signer and G.~Zanderighi,
%``Effective theory approach to unstable particle production,''
Phys.\ Rev.\ Lett.\  {\bf 93}, 011602 (2004)
[arXiv:hep-ph/0312331];
%%CITATION = HEP-PH 0312331;%%
%
M.~Beneke, A.~P.~Chapovsky, A.~Signer and G.~Zanderighi,
%``Effective theory calculation of resonant high-energy scattering,''
Nucl.\ Phys.\ B {\bf 686}, 205 (2004)
[arXiv:hep-ph/0401002].
%%CITATION = HEP-PH 0401002;%%


\bibitem{BaBarBook}
P.~F.~Harrison and H.~R.~Quinn  [BABAR Collaboration],
%``The BaBar physics book: Physics at an asymmetric B factory,''
SLAC-R-0504
%\href{http://www.slac.stanford.edu/spires/find/hep/www?r=slac-r-0504}{SPIRES
%entry} 
{\it Papers from Workshop on Physics at an Asymmetric B Factory (BaBar
  Collaboration Meeting), Rome, Italy, 11-14 Nov 1996, Princeton, NJ, 17-20 
Mar 1997, Orsay, France, 16-19 Jun 1997 and Pasadena, CA, 22-24 Sep 1997}


\bibitem{Korchemsky1}
G.~P.~Korchemsky and A.~V.~Radyushkin,
%``Infrared factorization, Wilson lines and the heavy quark limit,''
Phys.\ Lett.\ B {\bf 279}, 359 (1992)
[arXiv:hep-ph/9203222].
%%CITATION = HEP-PH 9203222;%%


\bibitem{Bauer1}
C.~W.~Bauer, D.~Pirjol and I.~W.~Stewart,
%``Soft-collinear factorization in effective field theory,''
Phys.\ Rev.\ D {\bf 65}, 054022 (2002)
[arXiv:hep-ph/0109045].
%%CITATION = HEP-PH 0109045;%%


\bibitem{ManoharQED}
A.~V.~Manohar and I.~W.~Stewart,
%``Logarithms of alpha in QED bound states from the renormalization group,''
Phys.\ Rev.\ Lett.\  {\bf 85}, 2248 (2000)
[arXiv:hep-ph/0004018].
%%CITATION = HEP-PH 0004018;%%

\bibitem{Fadin2}
V.~S.~Fadin, V.~A.~Khoze and A.~D.~Martin,
%``Interference radiative phenomena in the production of heavy unstable
%particles,''
Phys.\ Rev.\ D {\bf 49}, 2247 (1994).
%%CITATION = PHRVA,D49,2247;%%


\bibitem{Melnikov1}
K.~Melnikov and O.~I.~Yakovlev,
%``Top near threshold: All alpha-S corrections are trivial,''
Phys.\ Lett.\ B {\bf 324}, 217 (1994)
[arXiv:hep-ph/9302311].
%%CITATION = HEP-PH 9302311;%%

\bibitem{Jezabek1}
M.~Jezabek and J.~H.~K\"uhn,
%``QCD Corrections To Semileptonic Decays Of Heavy Quarks,''
Nucl.\ Phys.\ B {\bf 314}, 1 (1989).
%%CITATION = NUPHA,B314,1;%%

\bibitem{amis} A.V.~Manohar and I.W.~Stewart,
%``Renormalization group analysis of the QCD quark potential to order  v**2,''
Phys.\ Rev.\ D {\bf 62}, 014033 (2000)
[arXiv:hep-ph/9912226].
%%CITATION = HEP-PH 9912226;%%

\bibitem{amis2} A.V.~Manohar and I.W.~Stewart,
%``The QCD heavy-quark potential to order v**2: One loop matching
%conditions,'' 
Phys.\ Rev.\ D {\bf  62}, 074015 (2000)
[arXiv:hep-ph/0003032].
%%CITATION = HEP-PH 0003032;%%

\bibitem{Beneke2}
M.~Beneke and G.~Buchalla,
%``The $B_c$ Meson Lifetime,''
Phys.\ Rev.\ D {\bf 53}, 4991 (1996)
[arXiv:hep-ph/9601249].
%%CITATION = HEP-PH 9601249;%%


\bibitem{Blokland1}
I.~Blokland, A.~Czarnecki, M.~Slusarczyk and F.~Tkachov,
%``Heavy-to-light decays with a two-loop accuracy,''
Phys.\ Rev.\ Lett.\  {\bf 93}, 062001 (2004)
[arXiv:hep-ph/0403221].
%%CITATION = HEP-PH 0403221;%%


\bibitem{Denner1}
A.~Denner and T.~Sack,
%``The Top Width,''
Nucl.\ Phys.\ B {\bf 358}, 46 (1991).
%%CITATION = NUPHA,B358,46;%%


\bibitem{Modritsch1}
W.~M\"odritsch and W.~Kummer,
%``Relativistic and gauge independent off-shell corrections to the toponium
%decay width,''
Nucl.\ Phys.\ B {\bf 430}, 3 (1994).
%%CITATION = NUPHA,B430,3;%%


\bibitem{WiseTTbarSusy}
S.~f.~Su and M.~B.~Wise,
%``Supersymmetric correction to top quark pair production near threshold,''
Phys.\ Lett.\ B {\bf 510}, 205 (2001)
[arXiv:hep-ph/0104169].
%%CITATION = HEP-PH 0104169;%%


\bibitem{c12loop}
A.~Czarnecki and K.~Melnikov,
%``Two-loop QCD corrections to the heavy quark pair production cross  section
%in e+ e- annihilation near the threshold,''
Phys.\ Rev.\ Lett.\  {\bf 80}, 2531 (1998)
[arXiv:hep-ph/9712222];
%%CITATION = HEP-PH 9712222;%%
%
M.~Beneke, A.~Signer and V.~A.~Smirnov,
%``Two-loop correction to the leptonic decay of quarkonium,''
Phys.\ Rev.\ Lett.\  {\bf 80}, 2535 (1998)
[arXiv:hep-ph/9712302].
%%CITATION = HEP-PH 9712302;%%


\bibitem{Hoangc12loopQED}
A.~H.~Hoang,
%``Two-loop corrections to the electromagnetic vertex for energies close  to
%threshold,''
Phys.\ Rev.\ D {\bf 56}, 7276 (1997)
[arXiv:hep-ph/9703404].
%%CITATION = HEP-PH 9703404;%%


\bibitem{Bauer2}
C.~W.~Bauer, A.~F.~Falk and M.~E.~Luke,
%``Resumming Phase Space Logarithms in Inclusive Semileptonic B Decays,''
Phys.\ Rev.\ D {\bf 54}, 2097 (1996)
[arXiv:hep-ph/9604290].
%%CITATION = HEP-PH 9604290;%%


\bibitem{HoangTeubner}
A.~H.~Hoang and T.~Teubner,
%``Top quark pair production at threshold: Complete  next-to-next-to-leading
%order relativistic corrections,''
Phys.\ Rev.\ D {\bf 58}, 114023 (1998)
[arXiv:hep-ph/9801397].
%%CITATION = HEP-PH 9801397;%%

\bibitem{Hoangupsilon}
A.~H.~Hoang, Z.~Ligeti and A.~V.~Manohar,
%``B decay and the Upsilon mass,''
Phys.\ Rev.\ Lett.\  {\bf 82}, 277 (1999)
[arXiv:hep-ph/9809423];
%%CITATION = HEP-PH 9809423;%%
%
A.~H.~Hoang, Z.~Ligeti and A.~V.~Manohar,
%``B decays in the Upsilon expansion,''
Phys.\ Rev.\ D {\bf 59}, 074017 (1999)
[arXiv:hep-ph/9811239].
%%CITATION = HEP-PH 9811239;%%




\end{thebibliography}
\end{document}